\documentclass[10pt,a4paper]{article}
\usepackage{graphicx}
\usepackage{subfigure}
\usepackage{amsmath,amsthm}
\usepackage{amssymb,amsfonts}
\usepackage{authblk}
\usepackage{url}
\usepackage{hyperref}
\usepackage{textgreek}

\newtheorem{definition}{\textbf{Definition}}
\newtheorem{theorem}{\textbf{Theorem}}
\newtheorem{corollary}{\textbf{Corollary}}

\title{System Identification with Copula Entropy}
\author{Jian MA\thanks{Email: majian@hitachi.cn}}
\affil{Hitachi China Research Laboratory}
\date{}

\begin{document}

\maketitle

\begin{abstract}
	\noindent
	Identifying differential equation governing dynamical system is an important problem with wide applications. Copula Entropy (CE) is a mathematical concept for measuring statistical independence in information theory. In this paper we propose a method for identifying differential equation of dynamical systems with CE. The problem is considered as a variable selection problem and solved with the previously proposed CE-based method for variable selection. The proposed method is composed of two components: the difference operator and the CE estimator. Since both components can be done non-parametrically, the proposed method is therefore model-free and hyperparameter-free. The simulation experiment with the 3D Lorenz system verified the effectiveness of the proposed method.
\end{abstract}
{\bf Keywords:} {Copula Entropy; System Identification; Dynamical Systems}

\section{Introduction}
Differential equation is a main tool for modelling dynamical systems and has very wide applications in diverse fields. Discovering differential equation from data is a important problem in dynamical systems and is becoming popular recently. Many different approaches have been proposed to tackle this problem \cite{Kaptanoglu2023}.

Such equation discovery problem is usually considered as a regression problem in which the regression function between the states of systems and the derivatives of the system states is identified from data. Given a dynamical system that has mathematical model as
\begin{equation}
	\frac{dx_i}{dt}=f(\mathbf{x},t),
	\label{eq:de}
\end{equation}
where $x_i, i=1,\ldots,N$ are system states. The equation discovery is to identify the function $f$. Many regressions models have been applied to this problem, such as Gaussian processs \cite{Raissi2017}, SINDy with sparsity assumption \cite{Brunton2016}. 

In this paper, we propose to use copula entropy (CE) to identify the regression function in differential equations. CE is a mathematical concept for measuring statistical independence. It is defined with the copula theory by Ma and Sun \cite{Ma2011} and has many properties, such as non-positive, multivariate, invariant to monotonic transformation, equivalent to correlation coefficient in Gaussian cases. A non-parametric estimator of CE was also proposed \cite{Ma2011}.

Recently, CE was proposed to solve the variable selection problem \cite{Ma2021} and shown to be advantagous than the tradition similar methods, such as AIC, Lasso, etc. The CE-based variable selection method has been applied to many other fields, such as hydrology \cite{Chen2013}, medicine \cite{Mesiar2021}, manufacturing \cite{Sun2021}, reliability \cite{Sun2019}, energy \cite{Liu2022}, etc.

We propose a method for equation discovery based on the CE-based variable selection method. It is to estimate the CE values between the derivatives of system states and many candidate covariates derived from system states and then select those covariates associated with high CE values for the identified function $f$. The proposed method is composed of two components: the difference operator for calculating the derivatives of system states and the non-parametric CE estimator. Since both components can be done without any assumption on the underlying system, the proposed method is therefore model-free. In this paper we will verify the proposed method on the 3D Lorenz system, in which both first-order and second-order covariates are included in its system equations.

This paper is organized as follows: the theory and estimation of CE is introduce in Section \ref{sec:ce}, the proposed method is presented in Section \ref{sec:method}, the simulation experiment is given in Section \ref{sec:sim}, followed by some discussion in Section \ref{sec:dis}, finally Section \ref{sec:con} concludes the paper.

\section{Copula Entropy}
\label{sec:ce}
\subsection{Theory of CE}
Copula theory unifies representation of multivariate dependence \cite{nelsen2007,joe2014}. According to Sklar theorem \cite{Sklar1959}, a multivariate joint density function can be represented as a product of its marginals and copula density which represents dependence structure among random variables. Please refer to \cite{Ma2011} for notations.

With copula density, Ma and Sun \cite{Ma2011} defined a new mathematical concept, called Copula Entropy, as follows:
\begin{definition}[Copula Entropy]
	Let $\mathbf{X}$ be random variables with marginals $\mathbf{u}$ and copula density $c(\mathbf{u})$. The CE of $\mathbf{X}$ is defined as
	\begin{equation}
		H_c(\mathbf{x})=-\int_{u}{c(\mathbf{u})\log c(\mathbf{u})d\mathbf{u}}.
	\end{equation}
\end{definition}
In information theory, MI is a fundamental concept different from entropy \cite{infobook}. Ma and Sun \cite{Ma2011} proved that MI is essentially negative CE, as follows:
\begin{theorem}
	MI of random variables is equivalent to negative CE:
	\begin{equation}
		I(\mathbf{x})=-H_c(\mathbf{x}),
	\end{equation}
	where $I$ denotes MI.
	\label{theorem1}
\end{theorem}
Theorem \ref{theorem1} has a simple proof \cite{Ma2011} and an instant corollary on the relationship between the information contained in joint density function, marginals, and copula density.
\begin{corollary}
	The joint entropy of random variables equals to the sum of the entropies of each variable and the CE of the random variables:
	\begin{equation}
		H(\mathbf{x})=\sum_{i}{H_i(x_i)}+H_c(\mathbf{x}),	
	\end{equation}
	where $H$ denotes entropy.
\end{corollary}
The above worthy-a-thousand-word results cast insight into the relationship between MI and copula and therefore build a bridge between information theory and copula theory.

\subsection{Estimating CE}
MI, as a fundamental concept in information theory, has wide applications in different fields. However, estimating it has been notoriously difficult. Under the blessing of Theorem \ref{theorem1}, Ma and Sun \cite{Ma2011} proposed a simple and elegant nonparametric method for estimating CE(MI) from data, which is composed of two steps:
\begin{enumerate}
	\item estimating Empirical Copula Density (ECD);
	\item estimating CE from the estimated ECD. 
\end{enumerate}
Given a sample $\{X_1,\ldots,X_T\}$ i.i.d. generated from random variables $\mathbf{X}=\{X_1,\ldots,X_N\}$, one can easily derive ECD using empirical functions:
\begin{equation}
	F_i(x_i)=\frac{1}{T}\sum_{t=1}^{T}{I(X_t^i\leq x_t^i)},
\end{equation}
where $i=1,\ldots,N$ and $I$ denotes indicator function. Once ECD is estimated, estimating CE becomes a problem of entropy estimation which can be tackled with many existing methods. The k-Nearest Neighbor method \cite{Kraskov2004} was suggested in \cite{Ma2011} for such entropy estimation. In this way, a nonparametric method for estimating CE is derived.

\section{Proposed Method}
\label{sec:method}
In this section we propose a method for identifying dynamical system based on CE. Identifying dynamical system is treated as a problem of variable selection in which the covariates related to the derivatives of the system are selected. In this way, the CE-based method for variable selection can be applied to solve this problem. Particularly, the estimated CEs between system variables and derivatives are considered as criteria for variable selection.

The proposed method is composed of three steps:
\begin{enumerate}
	\item calculating the derivative of system variables with differential operator;
	\item estimating the CEs between the calculated derivatives and the covariates of the system;
	\item selecting the covariates with high CE value for each derivatives.
\end{enumerate}
With a time series of system variables, the derivative of them can be derived with difference operator. Given a pair of variable $X,Y$, the derivative of $y$ with respect to $x$ can be simply calculated as follows:
\begin{equation}
	\frac{dx}{dt}{\lvert}_{t=t_0} = \frac{x_{t_1}-x_{t_0}}{{t_1}-{t_0}}.
	\label{eq:diff}
\end{equation}

For the second step of the proposed method, the non-parametric estimator of CE can be used.

Since the difference operator and the CE estimator are both non-parametric, the proposed method is therefore model-free and can be applied to any system without any assumptions.

\section{Simulations}
\label{sec:sim}
\subsection{Experiments}
We conducted simulation experiments to verify the effectiveness of the proposed method. In the simulation, the 3D Lorenz system  \cite{Lorenz1963} is used to simulate time series data.

\paragraph{Lorenz system} The mathematical equations of the 3D Lorenz system are as follows:
\begin{equation}
\begin{aligned}
\frac{dx}{dt} &= \sigma (y-x),\\
\frac{dy}{dt} &= \rho x - y - xz, \\
\frac{dz}{dt}&= -\beta z + xy,\\
\end{aligned}
\label{eq:lorenz}
\end{equation}
where $\sigma,\rho,\beta$ are the parameters of the system for Prandtl number, Rayleigh number, and geometric factor. 

In the experiment, the time series data is first simulated from the 3D Lorenz system. Then the derivatives of $x,y,z$ with respect to $t$ are calculated from the simulate data based on \eqref{eq:diff}. Then the CE between the calculated derivatives and the covariates of the system equation are estimated from the simulated data using the non-parametric CE estimator \cite{Ma2011}. Here, six covariates are considered: $x,y,z,xy,xz,yz$ as candidates.

Two packages in \textsf{R} are used for the implementation of the experiments: the \texttt{nonlinearTseries} package for simulating the 3D Lorenz system and the \texttt{copent} package for estimating CE from data. In the simulation experiment, we set $\sigma = 10, \beta = 8/3, \rho = 28$ respectively. The starting points of the simulated time series is randomly generated. The time horizon is 30 and the sample rate is 100, which leads to 3000 samples for our experiment. The default values of the hyperparameters of the implemented CE estimator were used in the experiments.

\subsection{Results}
\begin{figure}
	\centering
	\includegraphics[width=\textwidth]{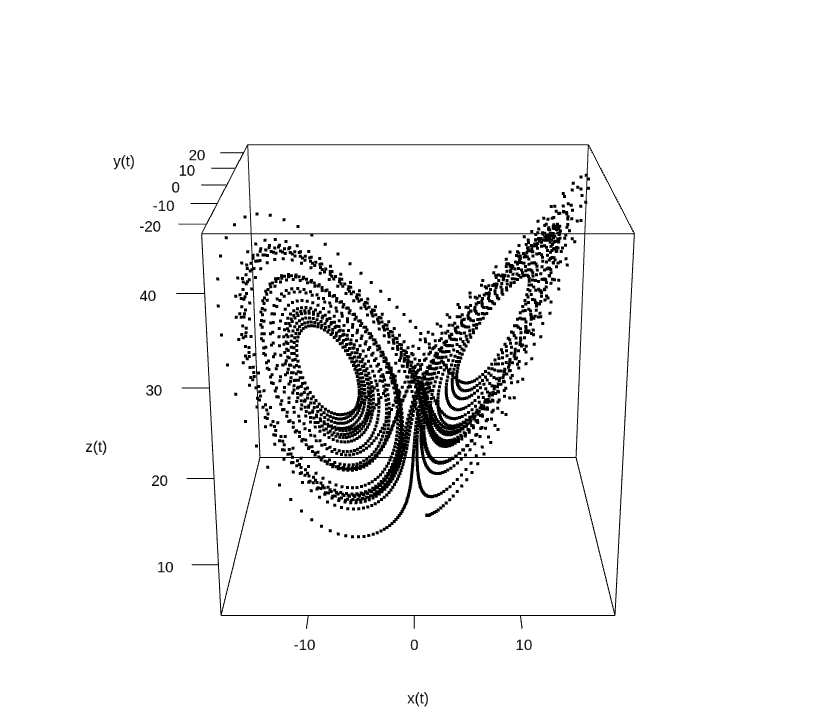}
	\caption{3D plot of the data simulated from the 3D Lorenz system.}
\label{fig:sim3d}
\end{figure}

\begin{figure}
	\centering
	\includegraphics[width=\textwidth]{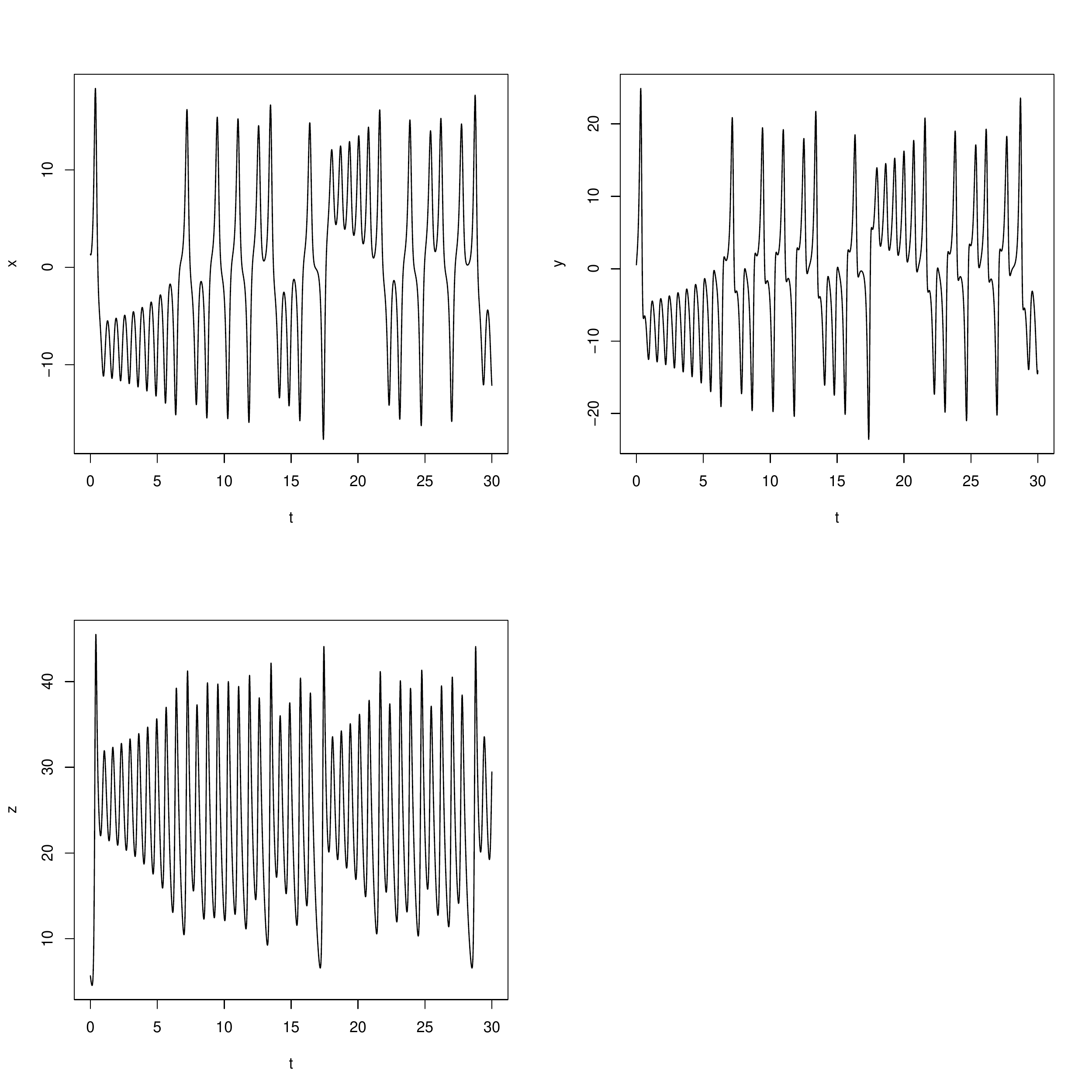}
	\caption{2D plots of the data simulated from the 3D Lorenz system.}
	\label{fig:sim2d}
\end{figure}

The simulated data from the 3D Lorenz system are shown in Figure \ref{fig:sim3d} and Figure \ref{fig:sim2d}.

\begin{figure}
	\centering
	\includegraphics[width=\textwidth]{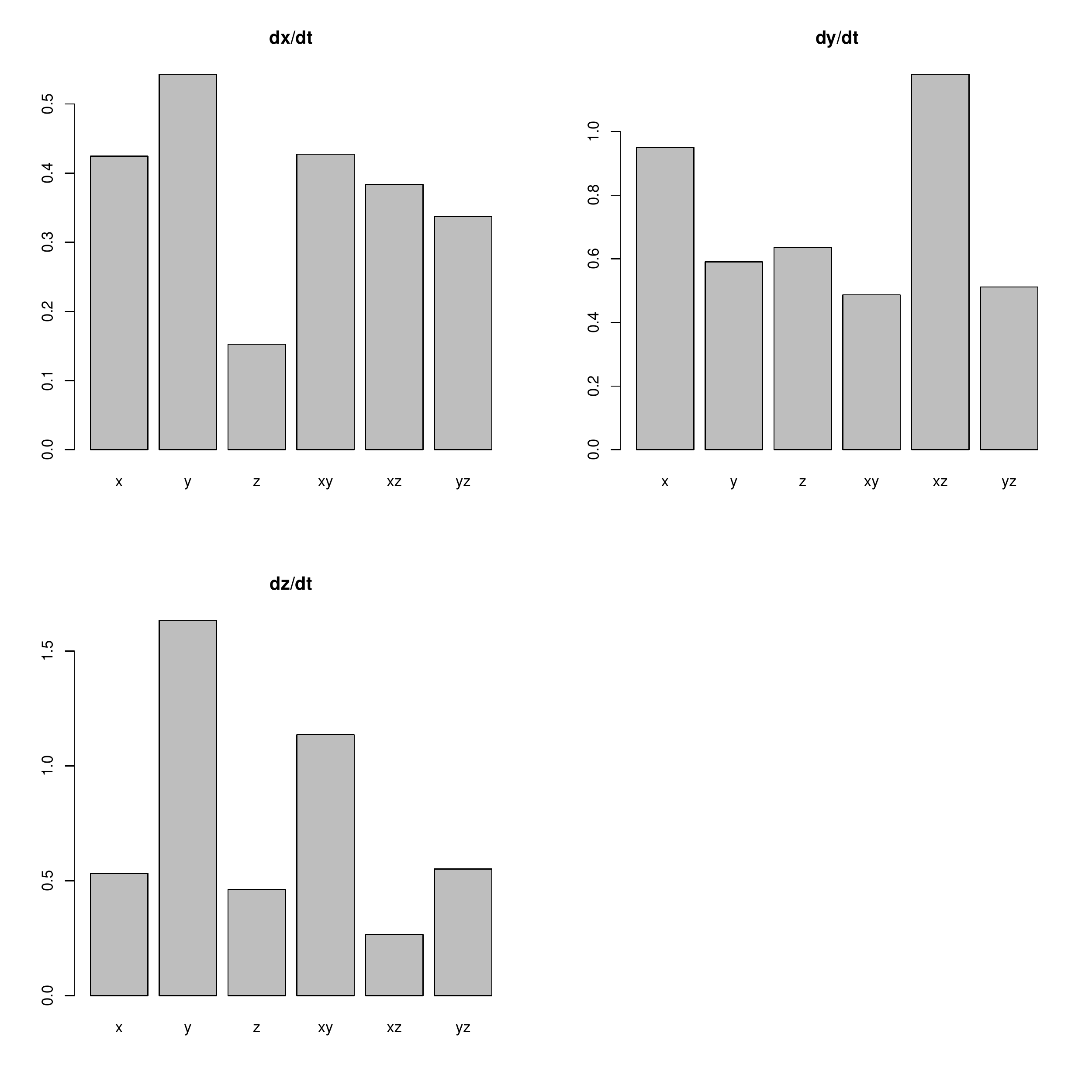}
	\caption{Estimated CEs between the candidate covariates and the derivatives of the states of the Lorenz system.}
	\label{fig:cexyz}
\end{figure}

After calculating the derivatives from the simulated data, the CEs between the derivatives and the covariates are estimated as shown in Figure \ref{fig:cexyz}. It can be learned from it that: 1) for the derivative of $x$, the covariates $x,y$ has large CE value while that of $z$ has small CE value; 2) for the derivative of $y$, the covariates $x,xz$ have large CE value; 3) for the derivative of $z$, the covariates $y,xy$ have large CE values. The large CE values means the relationships between the derivatives and the covariates.

Compared the estimation results with the system equation \eqref{eq:lorenz}, one can learn that the proposed method successfully identified the relationships between: the derivative of $x$ and $x,y$, the derivative of $y$ and $x,xz$, and the derivative of $z$ and $xy$. The two second-order covariates in \eqref{eq:lorenz} are both identified from the simulated data.

\section{Discussion}
\label{sec:dis}
In this paper, we proposed a method for system idenfication based on CE. the identification problem is treated as variable selection which has been solved with CE. Since CE can be estimated non-parametrically, the proposed method is also non-parametric and model-free. In practice, almost no hyper-parameter need to be tuned in estimating CE, which makes our method advantagous than the traditional methods that can only derive the results after hyperparameters-tuning.

CE is a rigorously defined mathematical concept in information theory, which makes it can be applied to not only statistical problems but also deterministic systems. It can measure the static relationships governed by system equations. It can also measure the dynamic relationships in the systems as information flow estimated with the CE-based estimator of transfer entropy \cite{Ma2019}, even in the systems with time delay \cite{Ma2023}.

In our experiment, only the first-order covariates $x,y,z$ and the second-order covariates $xy,xz,yz$ are considered due to the already known system equation. It can be easily extended to the cases where the system equation is unknown by considering more high-order covariates.

When selecting covariates based on the estimated CE values, the interrelationship between covariates should be considered. For example, the covariates related to the derivative of $x$ in our experiments include not only $x,y$ but $xy,xz,yz$ which are associated with high CE values. Such selection should be done carefully by considering these interrelationships between covariates.

\section{Conclusions}
\label{sec:con}
In this paper we propose a method for identifying differential equation of dynamical systems with CE. The problem is considered as a variable selection problem and solved with the previously proposed CE-based method. The proposed method is composed of two components: the difference operator and the CE estimator. Since both components can be done non-parametrically, the proposed method is therefore model-free and hyperparameter-free. The simulation experiment with the 3D Lorenz system verified the effectiveness of the proposed method.

\appendix
\section{Code}
The codes are available at \url{https://github.com/majianthu/sysid}.

\bibliographystyle{unsrt}
\bibliography{sysid}

\end{document}